\documentstyle[amssymb,psfig]{mn}

\newif\ifAMStwofonts
\ifoldfss

  \ifCUPmtlplainloaded \else
    \NewTextAlphabet{textbfit} {cmbxti10} {}
    \NewTextAlphabet{textbfss} {cmssbx10} {}
    \NewMathAlphabet{mathbfit} {cmbxti10} {} 
    \NewMathAlphabet{mathbfss} {cmssbx10} {} 
  \fi
  
\ifAMStwofonts
  
\ifCUPmtlplainloaded \else
      \NewSymbolFont{upmath} {eurm10}
      \NewSymbolFont{AMSa} {msam10}
      \NewMathSymbol{\upi}     {0}{upmath}{19}
      \NewMathSymbol{\umu}     {0}{upmath}{16}
      \NewMathSymbol{\upartial}{0}{upmath}{40}
      \NewMathSymbol{\leqslant}{3}{AMSa}{36}
      \NewMathSymbol{\geqslant}{3}{AMSa}{3E}

    \fi
    
\fi
\fi 
\ifnfssone
  
\newmathalphabet{\mathit}
  \addtoversion{normal}{\mathit}{cmr}{m}{it}
  \addtoversion{bold}{\mathit}{cmr}{bx}{it}

  \newmathalphabet{\mathbfit} 
  \addtoversion{normal}{\mathbfit}{cmr}{bx}{it}
  \addtoversion{bold}{\mathbfit}{cmr}{bx}{it}
  
\newmathalphabet{\mathbfss} 
  \addtoversion{normal}{\mathbfss}{cmss}{bx}{n}
  \addtoversion{bold}{\mathbfss}{cmss}{bx}{n}
  
\ifAMStwofonts
    
\ifCUPmtlplainloaded \else
      
      \UseAMStwoboldmath
      
\makeatletter
      \new@mathgroup\upmath@group
      \define@mathgroup\mv@normal\upmath@group{eur}{m}{n}
      \define@mathgroup\mv@bold\upmath@group{eur}{b}{n}
      \edef\UPM{\hexnumber\upmath@group}
      
\new@mathgroup\amsa@group
      \define@mathgroup\mv@normal\amsa@group{msa}{m}{n}
      \define@mathgroup\mv@bold\amsa@group{msa}{m}{n}
      \edef\AMSa{\hexnumber\amsa@group}
      \makeatother
      
\mathchardef\upi="0\UPM19
      \mathchardef\umu="0\UPM16
      \mathchardef\upartial="0\UPM40
      \mathchardef\leqslant="3\AMSa36
      \mathchardef\geqslant="3\AMSa3E

    \fi
  \fi
  
\fi 
\ifnfsstwo

  \DeclareMathAlphabet{\mathbfit}{OT1}{cmr}{bx}{it}
  \SetMathAlphabet\mathbfit{bold}{OT1}{cmr}{bx}{it}
  \DeclareMathAlphabet{\mathbfss}{OT1}{cmss}{bx}{n}
  \SetMathAlphabet\mathbfss{bold}{OT1}{cmss}{bx}{n}
  
\ifAMStwofonts
    
\ifCUPmtlplainloaded \else
      \DeclareSymbolFont{UPM}{U}{eur}{m}{n}
      \SetSymbolFont{UPM}{bold}{U}{eur}{b}{n}
      \DeclareSymbolFont{AMSa}{U}{msa}{m}{n}
      \DeclareMathSymbol{\upi}{0}{UPM}{"19}
      \DeclareMathSymbol{\umu}{0}{UPM}{"16}
      \DeclareMathSymbol{\upartial}{0}{UPM}{"40}
      \DeclareMathSymbol{\leqslant}{3}{AMSa}{"36}
      \DeclareMathSymbol{\geqslant}{3}{AMSa}{"3E}

    \fi
  \fi
  
\fi 
\ifCUPmtlplainloaded \else
  \ifAMStwofonts \else 
    
\def\upi{\pi}
\def\umu{\mu}
\def\upartial{\partial}
  \fi
\fi
\pubyear{1994}

\title{Small-scale CMB polarization anisotropies due to tangled primordial
magnetic fields}
\author[Kandaswamy Subramanian, T. R. Seshadri and John. D. Barrow]
{Kandaswamy Subramanian$^{1}$, T. R. Seshadri$^{2}$ and John. D. Barrow$^{3}$ \\
$^{1}$Inter University Centre for Astronomy and Astrophysics,
Post Bag 4, Ganeshkhind, Pune 411 007, India\\
$^{2}$ Department of Physics and Astrophysics, University of Delhi,
Delhi 110 007, India \\
$^{3}$DAMTP, Centre for Mathematical Sciences, Cambridge University,
Wilberforce Road, Cambridge CB3 0WA, UK}
\date{}
\begin{document}
\maketitle

\begin{abstract}
Tangled, primordial cosmic magnetic fields create small rotational
velocity perturbations on the last scattering surface (LSS) of
the cosmic microwave background radiation (CMBR). Such perturbations 
can contribute significantly to the CMBR temperature and polarization 
anisotropies at large $l>1000$ or so, like the excess power detected 
by the CBI experiment. The magnetic contribution can be distinguished 
from most conventional signals, as they lead to CMBR polarization 
dominated by the odd parity, B-type signal. Experiments like DASI and 
WMAP have detected evidence for CMBR polarization at small $l$.
Many experiments will also probe the large $l$ regime.
We therefore calculate the polarization signals due to 
primordial magnetic fields, for different spectra and different 
cosmological parameters. A scale-invariant spectrum of tangled fields 
which redshifts to a present 
value $B_{0}=3\times 10^{-9}$ Gauss, produces B-type polarization 
anisotropies of $\sim 0.3 - 0.4\mu K$ between $l\sim 1000-5000$. 
Larger signals result if the spectral index of magnetic tangles is 
steeper, $n>-3$. The peak of the signal shifts to larger $l$ for a 
lambda-dominated universe, or if the baryon density is larger.
The signal will also have non-Gaussian statistics. 
We also predict the much smaller E-type polarization, 
and T-E cross correlations for these models.
\end{abstract}

\label{firstpage}%

\section{Introduction}

The origin of large-scale cosmic magnetic fields remains an intriguing
question. It is quite likely that magnetic fields in astronomical objects,
like galaxies, grew by turbulent dynamo action on small seed magnetic fields
(cf. Ruzmaikin, Shukurov \& Sokoloff 1988; Beck et al 1996). However, this
idea is not without difficulties in view of the constraints implied by 
helicity conservation and the more rapid growth of small-scale magnetic 
fields (Cattaneo \& Vainshtein 1991; Kulsrud \& Anderson 1992; 
Gruzinov \& Diamond 1994; Subramanian 1998, 1999; 
Blackman \& Field 2000; Kleoorin {\it et al} 2000; 
Brandenburg 2001; Brandenburg \& Subramanian 2000;
Brandenburg, Dobler \& Subramanian 2002). 
Magnetic fields with larger coherence scales may also be present in 
clusters of galaxies (Clarke, Kronberg \& Bohringer 2001) and at high 
redshifts (Oren \& Wolfe 1995). Such large-scale coherent fields could 
present further problems for the dynamo paradigm. Alternatively, galactic 
or cluster fields could largely be a remnant of a primordial cosmological 
magnetic field (cf. Kulsrud 1990), although, as yet, there is no 
entirely compelling mechanism for producing the required field. 
They could be possibly generated during a period of inflation or at 
a phase transition, perhaps with an almost scale-invariant spectrum 
(Turner \& Widrow 1988; Ratra 1992; cf. Grasso \& Rubenstein 2001 
for a review). A primordial field, whose present-day strength is of order 
$10^{-9}$ Gauss, and is tangled on galactic scales, can also affect 
the process of galaxy formation (Rees \& Reinhardt 1972; Wasserman 1978; Kim,
Olinto \& Rosner 1996; Subramanian \& Barrow 1998a, SB98a hereafter). It is
of considerable interest, therefore, to find different ways of limiting or
detecting such primordial fields (see Kronberg 1994 and Grasso \& Rubenstein
2001, Widrow 2003 for reviews). 

The nearly isotropic nature of the CMBR already places a limit of 
several nano Gauss on the present strength of any {\it uniform} (spatially
homogeneous) component of the magnetic field (Barrow, Ferreira \& Silk
1997). Observations of CMBR anisotropies also provide potentially powerful 
constraints on tangled magnetic fields (Subramanian \& Barrow 1998b (SB98b),
2002 (SB02)). Such fields produce vortical perturbations, 
which are overdamped in the radiation era and can then survive Silk 
damping (Silk 1968) on scales much smaller than the compressional modes 
(Jedamzik, Katalinic \& Olinto 1998; SB98a). So their signal, if present, 
will be particularly evident at small angular scales below the 
conventional Silk damping scale or at multipoles  
$l > 1000$ or so (SB98b, SB02). Intriguingly, the CBI experiment 
identified significant excess power in the CMBR anisotropy 
spectrum up to $l = 3500$ (with 2 sigma limits of 
$ 14 - 31$ $\mu$K at $l > 2010$) (Mason {\it et al} 2002). 
We argued in SB02 that tangled magnetic fields 
which redshift to a present-day value of $B_{0} = 3 \times 10^{-9}$ Gauss 
(see below for a definition of $B_0$), can contribute a non-negligible 
fraction of this signal. Other possibilities include the Sunyaev-Zeldovich 
effect (Bond {\it et al} 2002; Komatsu \& Seljak 2002), primordial voids 
(cf. Griffiths {\it et al} 2002) or features in the primordial power 
spectrum (cf. Cooray \& Melchiorri 2002).
One needs to isolate and test for the possible 
magnetic field contribution. We point out here that this can be done 
by looking for the corresponding distinctive polarization signals.

Indeed, Seshadri and Subramanian (2001; Paper I)
showed that tangled magnetic fields create distinctive small-scale 
(large $l$) polarization anisotropy, dominated by the odd parity 
B-type signals. Mack {\it et al}.(2002; MKK02) gave estimates 
for such signals, but in the low $l < 500$ regime.
Recently the DASI experiment detected E-type polarization and both
DASI and WMAP have detected the T-E cross correlation,
albeit at smaller $l \sim 300$ (Kovac {\it et al} 2002; 
Kogut {\it et al} 2003). Several experiments to probe the polarization
anisotropy at the large $l$ regime are also underway or being planned
(VSA, ACBAR, ATCA, Polatron and Planck Surveyor). Motivated by these 
experiments, we compute the 
B-type polarization signals at larger values of $l$, for a wider variety 
of cosmological parameters and spectral indices, than were made in Paper I. 
In the next section, we first recapitulate the arguments 
of Paper I. We then present in Section 3, approximate
analytic estimates of the signals. The B-type anisotropy for various models, 
obtained from a detailed numerical integration is given in Section 4.
We also calculate there numerically the E-type polarization 
anisotropy and the T-E cross correlations for these models. 
Our predictions will allow comparison with future observations
and help in detecting or ruling out significant magnetic field
contributions to the signal on small angular scales.

\section{ The polarization anisotropy}

Polarization of the CMBR arises from the Thomson scattering of radiation
from free electrons, and is sourced by the quadrupole component of
the CMBR anisotropy. The evolution equations for the moments, $\Theta_l$,
$E_l$ and $B_l$, of the temperature
anisotropy ($\Delta T/T$), the electric (E-) type and the odd parity,
magnetic (B-) type polarization anisotropies, respectively,
for vector perturbations,
have been derived in detail by Hu \& White (1997a; HW97)
(see also Paper I and SB98b).
For vector perturbations, the B-type contribution dominates 
the polarization anisotropy (HW97). We therefore give details of 
its calculations and summarize the results for
the E-type contribution and the T-E cross correlations. 
The quadrupole anisotropy source term for polarization is given 
by $P(k,\tau) = [\Theta_2 - \sqrt{6} E_2]/10$.
Here $k$ is the co-moving wave number, $\tau $ the conformal time.
One can analytically estimate $P$ using the tight-coupling
approximation, $ k L_{\gamma}(\tau) \ll 1$, 
where $L_{\gamma}(\tau)$ is the co-moving, photon mean
free path. First, to leading order in this approximation,
we have zero quadrupoles, and  a dipole $\Theta_1 = v_B$,
where $v_B(k,\tau )$ is the magnitude of the rotational component 
of the fluid velocity $v_{i}^{B}$, in Fourier space. 
However, to the next order the quadrupole is
not zero. It is generated from the dipole at the `last but one'
scattering of the CMBR. From Eq. (60), (63) and (64) of HW97, we get 
$\Theta_2 = -4 E_2/\sqrt{6} =  4 k L_{\gamma} v_B/(3\sqrt{3})$ and hence
$ P = \Theta_2/4 = k L_{\gamma} v_B/(3\sqrt{3})$.   
Using this in Eq. (77) and (56) of HW97 gives an estimate for $B_l$
and the angular power spectra $C_l^{BB}$ due to B-type polarization
anisotropy (cf. Paper I),              
\begin{eqnarray}
C_{l}^{BB} &=&4\pi {\frac{(l-1)(l+2)}{l(l+1)}}
\int_{0}^{\infty }{\frac{k^{2}dk}{2\pi ^{2}}}\quad {\frac{%
l(l+1)}{2}}  \nonumber \\
\  &&\times <|\int_{0}^{\tau _{0}}d\tau g(\tau _{0},\tau )
({\frac{kL_\gamma(\tau)}{3}})v_B(k,\tau ) \nonumber \\ 
&&\times {\frac{j_{l}(k(\tau _{0}-\tau ))}{k(\tau _{0}-\tau )}}|^{2}>.  
\label{deldef}
\end{eqnarray}
Here $j_{l}(z)$ is the spherical Bessel function of order $l$,
and $\tau_0$ the present value of $\tau$.
The 'visibility function', $g(\tau _{0},\tau ),$ determines the 
probability that a photon reaches us at epoch $\tau _{0}$ if it 
was last scattered at the epoch $\tau $. We adopt a flat universe 
throughout, with a total matter density $\Omega _{m}$ and a non-zero 
cosmological constant density $\Omega _{\Lambda }=1-\Omega _{m}$ today.
We now briefly recall the arguments detailed in Paper I.

Firstly, we approximate the visibility function as a Gaussian: 
$g(\tau _{0},\tau)=(2\pi \sigma ^{2})^{-1/2}
\exp [-(\tau -\tau _{\ast })^{2}/(2\sigma ^{2})]$%
, where $\tau _{\ast }$ is the conformal epoch of ``last scattering'' and $%
\sigma $ measures the width of the LSS. 
To estimate these, we use the WMAP results (cf. Spergal {\it et al} 2003),
that the redshift of LSS $ z_\ast = 1089$ and its thickness (FWHM) 
$\Delta z = 194$. To convert redshift into conformal time we use, 
$\tau =6000h^{-1}((a+a_{eq})^{1/2}-a_{eq}^{1/2})/\Omega _{m}^{1/2}$, 
valid for a flat universe (cf. Hu \& White 1997b). 
Here, the expansion factor $%
a=(1+z)^{-1}$ and $a_{eq}=4.17\times 10^{-5}(\Omega_{m}h^{2})^{-1}$ 
($h$ is the Hubble constant in units of $100$ km s$^{-1}$ Mpc$^{-1}$).  
For the $\Lambda $-dominated model suggested by WMAP
results, with $\Omega_m =0.27$, $\Omega_\Lambda = 0.73$,
we get $\tau _{\ast }=201.4h^{-1}$ Mpc and $\sigma =11.5h^{-1}$ Mpc. 
For an $\Omega _{m}=1$ model, with the same baryon density,
(cf. $\Omega_b = 0.0224h^{-2}$), we use the expressions given 
in Hu and Sugiyama (1995) to estimate 
$\tau _{\ast }=131.0h^{-1}$ Mpc, and $\sigma =8.3h^{-1}$ Mpc.
We use these numbers in the numerical estimates below.

To evaluate $C_{l}^{BB}$, one needs to estimate $v_B$, 
the rotational velocity induced by magnetic inhomogeneities. 
The magnetic field is assumed to be initially a Gaussian random field. 
On galactic scales and above, the induced velocity is generally so small 
that it does not lead to any appreciable distortion of the initial field 
(Jedamzik, Katalinic \& Olinto 1998, SB98a). So,
the magnetic field simply redshifts away as ${\bf B}(%
{\bf x},t)={\bf b}_{0}({\bf x})/a^{2}$. The Lorentz force associated with
the tangled field is then ${\bf F}_{L}=({\bf \nabla }\times {\bf b}%
_{0})\times {\bf b}_{0}/(4\pi a^{5})$, which pushes the fluid and creates
rotational velocity perturbations. These can be estimated as in 
SB02 or Paper I, by using the Euler equation for the baryons. 
On scales larger than the photon mean-free-path at decoupling, 
where the viscous effect due to photons can be
treated in the diffusion approximation, this reads (SB02) 
\begin{equation}
\left( {\frac{4}{3}}\rho _{\gamma }+\rho _{b}\right) {\frac{\partial 
v_i^B}{\partial t}}+\left[ {\frac{\rho _{b}}{a}}{\frac{da}{dt}}+{\frac{%
k^{2}\eta }{a^{2}}}\right] v_i^B={\frac{P_{ij}F_{j}}{4\pi a^{5}}}.
\label{eulerk}
\end{equation}
Here, $\rho _{\gamma }$ is the photon density, $\rho _{b}$ the baryon
density, and $\eta =(4/15)\rho _{\gamma }l_{\gamma }$ the shear viscosity
coefficient associated with the damping due to photons, whose mean-free-path
is $l_{\gamma }=(n_{e}\sigma _{T})^{-1}\equiv L_{\gamma }a(t)$, where $n_{e}$
is the electron density and $\sigma _{T}$ the Thomson cross-section.
We have also ignored here a metric perturbation term which is subdominant at
large $l$ (cf. Paper I). 
For $z_\ast \sim 1089$, we get $L_\gamma(\tau_*) \sim 1.83 f_b^{-1}$ Mpc,
where $f_{b}=(\Omega_{b}h^{2}/0.0224)$ (the WMAP value). We have
defined the Fourier transforms of the magnetic field, by ${\bf b}_{0}({\bf x}%
)=\sum_{{\bf k}}{\bf b}({\bf k})\exp (i{\bf k}.{\bf x})$ and ${\bf F}({\bf k}%
)=\sum_{{\bf p}}[{\bf b}({\bf k}+{\bf p}).{\bf b}^{\ast }({\bf p})]{\bf p}-[%
{\bf k}.{\bf b}^{\ast }({\bf p})]{\bf b}({\bf k}+{\bf p})$. The projection tensor, 
$P_{ij}({\bf k})=[\delta _{ij}-k_{i}k_{j}/k^{2}]$ 
projects ${\bf F}$ onto its transverse components perpendicular to ${\bf k}$.

The comoving Silk damping scale at recombination, $L_{S}=k_{S}^{-1}\sim 10$
Mpc, separates scales on which the radiative viscosity is important ($%
kL_{S}\gg 1$) from those on which it is negligible ($kL_{S}\ll 1$). For $%
kL_{s}\ll 1$, the damping due to the photon viscosity can be neglected
compared to the Lorentz force. Assuming negligible initial rotational
velocity perturbation, we can integrate the baryon Euler equation to get
$v_i^B=G_{i} D$, where $G_{i}=3P_{ij}F_{j}/[16\pi \rho _{0}]$ 
and $D = \tau /(1+S_{\ast })$ (see SB02 or paper I). Here 
$\rho _{0}$ is the present-day value of $\rho _{\gamma }$, and $S_{\ast
}=(3\rho _{b}/4\rho _{\gamma })(\tau _{\ast })\sim 0.59 f_b$. 
For $kL_{s} \gg 1$, we use the terminal-velocity approximation, 
neglecting the inertial terms in the Euler
equation, to balance the Lorentz force by friction. This gives 
$v_i^B=G_{i}({\bf k}) D$, but with now $D = (5/k^2L_{\gamma})$, 
on scales where diffusion damping operates.
The transition Silk scale can also be estimated by equating $v_i^B$ in
the two cases, to give $k_{S}\sim \lbrack 5(1+S_{\ast })/(\tau L_{
\gamma }(\tau))]^{1/2}$.

To compute the $C_{l}^{BB}$s we need to specify the 
spectrum of the tangled magnetic field, say $M(k)$. We define, 
$<b_{i}({\bf k})b_{j}({\bf q}%
)>=\delta _{{\bf k},{\bf q}}P_{ij}({\bf k})M(k)$, where $\delta _{{\bf k},%
{\bf q}}$ is the Kronecker delta which is non-zero only for ${\bf k}={\bf q}$%
. This gives $<{\bf b}_{0}^{2}>=2\int (dk/k)\Delta _{b}^{2}(k)$, where $%
\Delta _{b}^{2}(k)=k^{3}M(k)/(2\pi ^{2})$ is the power per logarithmic
interval in $k$ space residing in magnetic tangles, and we replace the
summation over $k$ space by an integration. The ensemble average $<|v_B|^{2}>$%
, and hence the $C_{l}^{BB}$s, can be computed in terms of 
the magnetic spectrum $%
M(k)$. It is convenient to define a dimensionless spectrum, $m(k)=\Delta
_{b}^{2}(k)/(B_{0}^{2}/2)$, where $B_{0}$ is a fiducial constant magnetic
field. The Alfv\'{e}n velocity, $V_{A}$, for this fiducial field is, 
\begin{equation}
V_{A}={\frac{B_{0}}{(16\pi \rho _{0}/3)^{1/2}}}\approx 3.8\times
10^{-4}B_{-9},  \label{alfvel}
\end{equation}
where $B_{-9}\ \equiv (B_{0}/10^{-9}{\rm Gauss})$. We will also consider
as in SB02, power-law magnetic spectra, $M(k)=Ak^{n}$ cut-off at $k=k_{c}$, 
where $k_{c}$ is the Alfv\'{e}n-wave damping length-scale 
(Jedamzik, Katalinic \& Olinto, SB98a). We fix $A$ by demanding 
that the smoothed field strength over a
''galactic'' scale, $k_{G}=1h{\rm Mpc}^{-1}$, (using a sharp $k$-space
filter) is $B_{0}$, giving a dimensionless spectrum for $n>-3$ of 
\begin{equation}
m(k)=(n+3)(k/k_{G})^{3+n}.  
\label{powspec}
\end{equation}

\section{ Analytic estimates }

The dominant contributions to the integral over $\tau $ in Eq. (\ref{deldef}%
) come from a range $\sigma $ around the epoch $\tau =\tau _{\ast }$.
Furthermore, $j_{l}(k(\tau _{0}-\tau ))$ picks out $(k,\tau )$ values in the
integrand which have $k(\tau _{0}-\tau )\sim l.$ Thus, following the
arguments detailed in Paper I and SB02, for $k\sigma \ll 1$ we get the 
analytical estimate, $l(l+1)C_{l}^{BB}/(2\pi )\approx 
(kL_\gamma(\tau{\ast})/3)^2 (\pi /4)\Delta _{v}^{2}(k,\tau _{\ast
})|_{k=l/R_{\ast }}$. Here, $\Delta _{v}^{2}=k^{3}<|v_B(k,\tau _{\ast
})|^{2}>/(2\pi ^{2})$ is the power per unit logarithmic interval of $k$,
residing in the {\it net} rotational velocity perturbation, 
and $R_{\ast }=\tau_{0}-\tau _{\ast }$. In the opposite limit, 
$k\sigma \gg 1$, we get $%
l(l+1)C_{l}^{BB}/(2\pi )\approx (kL_\gamma(\tau{\ast})/3)^2 
(\sqrt{\pi }/4)(\Delta _{v}^{2}(k,\tau _{\ast
})/(k\sigma )|_{k=l/R_{\ast }}$. At small wavelengths, $C_{l}^{BB}$ 
is suppressed by a $1/k\sigma $ factor due to the finite thickness of the LSS.
Further in both cases, the polarization
anisotropy, $\Delta T_P^{BB}(l) \approx (kL_{\gamma}(\tau_*)/ 3)
\times \Delta T(l)$, where, $\Delta T(l)$ is the temperature anisotropy
computed in SB02. We can now put together the above results to derive 
approximate analytic estimates for the CMBR polarization anisotropy induced by 
tangled magnetic fields. As a measure of the anisotropy we define 
the quantity $\Delta T_P^{BB}(l)\equiv \lbrack
l(l+1)C_{l}^{BB}/2\pi ]^{1/2}T_{0}$, where $T_{0}=2.728$ K is the CMBR
temperature. On large scales, such that $kL_{s}<1$ and $k\sigma <1$, the
resulting CMBR anisotropy is (see Paper I) 
\begin{eqnarray}
\Delta T_{P}^{BB}(l) &=&T_{0}({\frac{\pi }{32}})^{1/2}I(k)
{\frac{k^2 L_\gamma(\tau_{\ast}) V_{A}^{2}\tau
_{\ast }}{3(1+S_{\ast })}}  \nonumber \\
\  &\approx &0.4\mu K\left( {\frac{B_{-9}}{3}}\right) ^{2}\left( {\frac{l}{%
1000}}\right)^2 I({\frac{l}{R_{\ast }}}).  \label{largT}
\end{eqnarray}
Here, $l=kR_{\ast }$ and we adopted the $\Lambda $-dominated model
favored by WMAP, with $\Omega _{\Lambda }=0.73$, $\Omega _{m}=0.27$,
$\Omega_{b}h^{2}=0.0224$ and $h=0.71$ 
(in Paper I, we used a purely matter-dominated $%
\Omega _{m}=1$ model). We also use the fit given by Hu \& White (1997b) 
to calculate 
$\tau_{0}=6000h^{-1}((1+a_{eq})^{1/2}-a_{eq}^{1/2})(1-0.0841 \ {\rm ln}
(\Omega_{m}))/\Omega _{m}^{1/2}$, valid for flat universe.
On scales where $kL_{S}>1$ and $k\sigma >1$, but $kL_{\gamma }(\tau _{\ast
})<1$, we get 
\begin{eqnarray}
\Delta T_{P}^{BB}(l) &=&T_{0}{\frac{\pi ^{1/4}}{\sqrt{32}}}I(k)
{\frac{5V_{A}^{2}}{3(k\sigma )^{1/2}}}  \nonumber \\
\  &\approx &1.2\mu K\left( {\frac{B_{-9}}{3}}\right) ^{2}\left( {\frac{l}{%
2000}}\right)^{-1/2}I({\frac{l}{R_{\ast }}}).  \label{smalT}
\end{eqnarray}
The function $I^{2}(k)$ in the Eqs.(\ref{largT}) and (\ref{smalT}) is a
dimensionless mode-coupling integral given in our previous 
papers (cf. Eq. (7) of Paper I). Analytic approximations to $I(k)$
exist for power-law spectra and for $k \ll k_c$ (as 
generally relevant even at high $l$; cf. Paper I and MKK02) For $n>-3/2$, 
\begin{equation}
I^{2}(k)={\frac{28}{15}}{\frac{(n+3)^{2}}{(3+2n)}}({\frac{k}{k_{G}}})^{3}({%
\frac{k_{c}}{k_{G}}})^{3+2n},  \label{mklar}
\end{equation}
dominated by the cut-off scale $k_c$. For $n<-3/2$ (SB02), 
\begin{equation}
I^{2}(k)={\frac{8}{3}}(n+3)({\frac{k}{k_{G}}})^{6+2n}  \label{mksmal}
\end{equation}
independent of $k_c$, where we neglect a subdominant term of order 
$(k_c/k)^{3+2n} \ll 1$; note that $k \ll k_c$.
A nearly scale-invariant spectrum, say with $n=-2.9$, then gives $%
\Delta T_P^{BB}(l)\sim 0.16\mu K(l/1000)^{2.1}$ for scales larger than 
the Silk scale, and $\Delta T_P^{BB}(l)\sim 0.51\mu K(l/2000)^{-0.4}$, 
for scales smaller than $L_{S}$ but larger than $L_{\gamma }$. 
Larger signals result for steeper spectra, $n>-2.9$ at the higher $l$ end.

\begin{figure}
\begin{picture}(240,240)
\psfig{figure=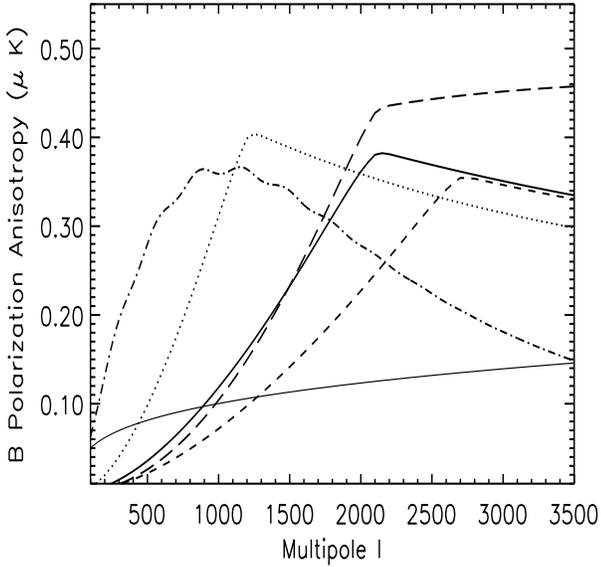,height=8.5cm,width=8.5cm,angle=0.0}
\end{picture}
\caption{$\Delta T_P^{BB}$ versus $l$ predictions for different cosmological 
models and magnetic power spectrum $M(k)\varpropto k^{n}$, for $B_{-9}=3$. 
The bold solid line is for a standard flat,  
$\Lambda $-dominated model, with $\Omega _{\Lambda }=0.73$, 
$\Omega _{m}=0.27$, $\Omega_{b}h^{2}=0.0224$, $h=0.71$ and almost scale 
invariant spectrum $n=-2.9$. The long dashed curve obtains
when one changes to $n=-2.5$, while the short dashed curve 
is for a larger baryon density $\Omega_{b}h^{2}=0.03$.
The dotted curve gives results for a $\Omega _{m}=1$ and 
$\Omega_{\Lambda }=0$ model, with $n=-2.9$. 
These curves show the build up of power in B-type polarization,
due to vortical perturbations from tangled magnetic fields which 
survive Silk damping at high $l \sim 1000-3500$. The eventual 
flattening or slow 
decline is due to the damping by photon viscosity, although this 
is only a mild effect as the magnetically sourced vortical mode 
is overdamped. We also show for qualitative comparison 
(dashed-dotted curve), the B-type polarization anisotropy due to 
gravitational lensing, in the canonical $\Lambda $-CDM model, computed 
using CMBFAST (Seljak \& Zaldarriaga 1996; Zaldarriaga \& Seljak 1998). 
The signal due to magnetic 
tangles dominate for $l$ larger than about $1000$. Finally, the thin
solid line gives the expected galactic foreground contribution
estimated by Prunet {\it et al} (1998), which is also smaller
than the predicted signals. }
\end{figure}

\section{Numerical results}

We now compute $\Delta T_P^{BB}(l)$ for the above spectra, 
by evaluating the $\tau $ and $k$ integrals in Eq.(%
\ref{deldef}) numerically. For this we first express 
Eq.\ (\ref{deldef}) explicitely in terms of the magnetic correlation 
function. We have for $l \gg 1$, 
$(\Delta T_P^{BB}(l)/ T_0)^2
= \int_{0}^{\infty }(dk/k) (l^2 k V_A^2 I(k) U(k)/ \sqrt{8})^2$, 
where
\begin{equation}
U(k) = \int_{0}^{\tau _{0}}d\tau \ g(\tau _{0},\tau )
{\frac{kL_\gamma(\tau)}{3}} D(k,\tau )
{\frac{j_{l}(k(\tau _{0}-\tau ))}{k(\tau _{0}-\tau )}}.  
\label{deldefn}
\end{equation}
In doing the above integral numerically, we 
retain the analytic approximations to $I(k)$
and $D$ with a transition between the limiting forms of $D$, at
wavenumber $k_{S}$. (The function $j_l$ is treated as in
codes like CMBFAST). The numerically evaluated results are shown 
in Figures 1 and 2.

We see that for $B_{0}\sim 3\times 10^{-9}G$, this leads to a predicted
RMS B-type polarization anisotropy in the CMBR of order $0.3-0.4\mu K$ 
for $1000<l<5000$, for a nearly scale-invariant power law spectra with 
$n=-2.9$. Larger signals result, at the high $l$ end, 
even for a moderately steeper spectral index, with $n=-2.5$, 
which has more power on small-scales 
(compare the long dashed and solid curves). 
Much larger signals result from even steeper spectra with $n > -2.5$.
However, these cases are probably ruled out for $B_{-9}\sim 3$, as they
lead to an over production of gravitational waves 
(Caprini \& Durrer 2002). Hence we do not display the results
for such spectra explicitely.
The peak of the polarization signal shifts to
larger values of $l$ with increasing $\Lambda $ 
(compare the solid and dotted curves), 
due to an increase in $R_{\ast }$. This also
happens for larger baryon density (compare the solid and 
short dashed curves), due to the decrease in $L_{\gamma }$ and hence 
the damping effects of radiative viscosity. 
The $l$-dependence of $\Delta T_P^{BB}(l)$ got by analytic approximation 
matches very well with that obtained by numerical integration. 
The amplitude is, however, somewhat overestimated by the analytic
treatment; by a factor of about $1.2$ and $1.6$ respectively,
for the regimes above and below the Silk scale.
So one can indeed get a reasonable
idea of the signals from the analytics, but for more accurate
amplitudes, one needs the numerical integration done above.

We have also numerically computed the E-type polarization ($C_l^{EE}$) 
as well as the T-E cross correlation ($C_l^{TE}$) for the above models. 
For this we use Eqns. (77), (79)  and (56) given in HW97 
and evaluate the $k$ and $\tau$ integrals numerically, adopting the 
same analytic approximations for $I(k)$ and $D$ given above.
In calculating the T-E signal, we have included a small 
polarization contribution to T as well, which contributes negligibly 
to T, but significantly to $C_l^{TE}$. 
The cross correlation due to the vortical modes induced
by magnetic tangles has a negative sign and so we define 
the corresponding effective 'temperature' after taking
a modulus of $C_{l}^{TE}$.
We show in Figure 2 the resulting E and T-E anisotropies,
for the standard $\Lambda$-dominated model. For comparison, we have
also give the temperature (T) and B-type polarization anisotropies 
(for T we have corrected a normalisation error made in SB02). 
The E-type polarization has a peak value of 
$\sim 0.1 \mu $ K and so is much smaller that the B-type signal,
as expected for vector perturbations (cf. HW97).
The T-E cross correlation is $\sim 0.1-0.2 \mu K$,
for $ l> 1000$. However, both E and T-E power are subdominant
to that produced by the standard scalar perturbations.

\begin{figure}
\begin{picture}(240,240)
\psfig{figure=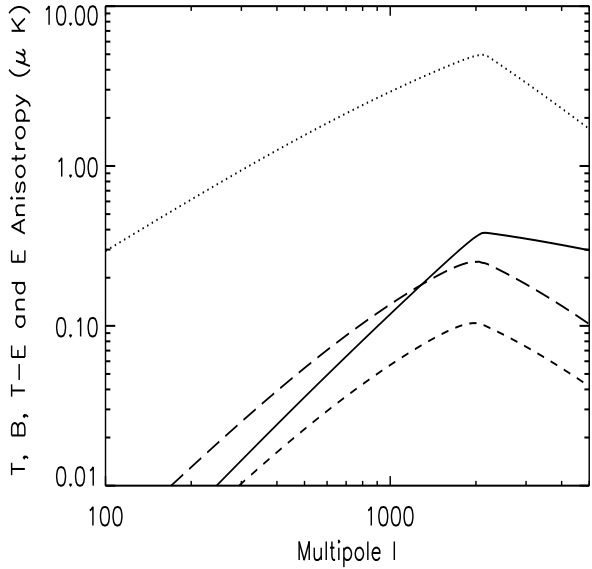,height=8.5cm,width=8.5cm,angle=0.0}
\end{picture}
\caption{The predicted anisotropy in temperature (dotted line),
B-type polarization (solid line), E-type polarization
(short dashed line) and T-E cross correlation (long dashed line)
up to large $ l\sim 5000$ for the standard $\Lambda$-CDM model,
due to magnetic tangles with a nearly scale invariant spectrum.
}
\end{figure} 

\section{Discussion}

We have re-examined the small angular scale polarization
anisotropy induced by tangled magnetic fields, for different cosmological 
parameters and spectral indices. A major motivation arises from 
ongoing and future experiments, which probe this large $l$ regime. 
Tangled magnetic fields generate vortical perturbations, 
which lead to a distinctive polarization anisotropy
dominated by a B-type contribution. 
Vortical perturbations survive Silk damping on much 
smaller scales than do compressional modes and
their damping due to the finite thickness of the LSS is also milder.
By contrast, in the standard non-magnetic models
the polarization anisotropy is dominated by E-type contributions.
A scale-invariant spectrum of tangled fields which redshifts to a 
present value of $B_{0}=3\times 10^{-9}$ Gauss, produces B-type 
polarization anisotropies of order $\sim 0.3-0.4 \mu K$ between 
$l\sim 1000-5000$. Larger signals are produced for steeper spectra 
with $ n > -3$. These signals are also larger than the small B-type
polarization induced by gravitational lensing at
$l> 1000$ or the expected galactic foreground contribution.
The peak of the signal shifts to larger $l$, in a $\Lambda $-dominated 
universe, or in a universe with larger $\Omega_b$. 
Our results complement the small $l$ ($l < 500$), and 
purely analytical estimates of MKK02.
The polarization anisotropy peaks or troughs could be much larger, 
because the non-linear dependence of $C_{l}^{BB}$ on $M(k)$,  
implies non-Gaussian statistics for the anisotropies.
Clearly, with the sub-micro-Kelvin sensitivities expected
from experiments like the Planck Surveyor, these signals can be detected. 

An important contributor to small scale anisotropies, 
like that seen the CBI experiment, would be the Sunyaev-Zeldovich effect.
This signal can be isolated by its frequency dependence. Also, 
scattering in clusters produces a much smaller statistical polarization 
anisotropy compared to the magnetically-induced signals
(even individual clusters produce maximum signals of 
only $\sim 0.1 \mu$K (cf. Sazonov \& Sunyaev, 2000)).
The polarization induced by the SZ effect, primordial voids,
or features in the power spectrum are all
expected to be predominantly E-type, in contrast to predominantly
B-type signals predicted here. Primordial magnetic fields
can also lead to depolarization due to differential
Faraday rotation. This effect is only important 
at frequencies lower than about $16.4 {\rm GHz} (B_{-9}/3)^{1/2}$
(Kosowsky \& Loeb 1996; Harari, Hayward \& Zaldariaga 1997).
Additional B-type polarization can arise from tensor modes generated 
during inflation or those induced by the primordial magnetic fields
(cf. Durrer, Ferreira \& Kahniashvili 2000; MKK02; SB02); 
but these are expected to be important only at $l < 100$ or so.
Helicity in the magnetic spectrum can also leave
interesting signatures on the CMBR (Pogosian, Vachaspati \& 
Winitzki 2002).

In summary, sensitive observational searches for
B-type polarization anisotropies at large $l$
will allow us to detect or constrain primordial, tangled magnetic fields.
They will also tell us whether such fields are significant
contributors to the excess power at large $l$ detected
by the CBI experiment and help probe possible new physics in the 
early universe.

\section*{Acknowledgments}
We would like to thank Pedro Ferreira for discussions.

\label{lastpage}
\end{document}